\documentstyle[aps,floats,graphicx]{revtex}

\begin{document}

\newcommand{\bec}{\begin{center}}
\newcommand{\ec}{\end{center}}
\newcommand{\be}{\begin{equation}}
\newcommand{\ee}{\end{equation}}
\newcommand{\beqn}{\begin{eqnarray}}
\newcommand{\eeqn}{\end{eqnarray}}
\newcommand{\bet}{\begin{table}}
\newcommand{\ent}{\end{table}}
\newcommand{\bib}{\bibitem}

\wideabs{

\title{
Simulated ion-sputtering and Auger electron spectroscopy depth profiling study of intermixing
in Cu/Co
}

\author{M. Menyh\'ard, P. S\"ule}
  \address{Research Institute for Technical Physics and Materials Science,\\
Konkoly Thege u. 29-33, Budapest, Hungary,sule@mfa.kfki.hu,www.mfa.kfki.hu/$\sim$sule\\
}

\date{\today}

\maketitle

\begin{abstract}
The ion-bombardment induced evolution of intermixing is studied by molecular dynamics simulations
and by Auger electron spectroscopy depth profiling analysis (AESD) in Cu/Co multilayer.
It has been shown that
from AESD we can derive
the low-energy mixing rate and which can be compared with
the simulated values obtained by molecular dynamics (MD) simulations.
The overall agreement is reasonably good hence MD can hopefully be used to estimate
the rate of intermixing in various interface systems.


\end{abstract}
}

  Low-energy ion beams are commonly used in surface analysis and for film growth \cite{Donnelly,IAFG}.
  The use of ion-sputtering in the controllable production of nanostructures and self-assembled nanoppaterns  have also become one of the most important fields
in materials science \cite{IAFG,Chen,Persson,Timm,Facsko,Gomez}.

 The future of nanotechnology ultimately rests on the controllable fabrication, integration, and stability of nanoscale devices.
However, the understanding of the fundamental phenomena leading to the formation, stability, and morphological evolution of nanoscale features is lacking \cite{Schukin}.
As the dimensions of the surface features is reduced to the nanoscale, many classical macroscopic (continuum and mesoscopic) models for morphological evolution lose their validity.
Therefore the understanding of the driving forces and laws governing mass transport involved in the synthesis and organisation of nanoscale features in solid state materials
is inevitable.
Unfortunately the fundamental understanding and the nanoscale control of interdiffusion 
is not available yet \cite{Adamowich,Ladwig}.
In order to get more insights in the atomic relocation processes during postgrowth
low-energy ion-sputtering, it is important to measure and to calculate accurately the rate of
intermixing at the interface.

 In this Letter we will show that using a newly developed code for simulated ion-sputtering based on molecular dynamics
we are able to get mixing rates.
We will also show that it is possible to extract the mixing rate data from AESD.
It turns out that the agreement of the two methods is reasonably good.
In this way we also could check the reliability of molecular dynamics simulations.

  The Auger electron spectroscopy depth profiling analysis (AESD) \cite{Hofmann,Barna} measurements were carried out on a Cu/Co multilayer system. The sample was made by sputter deposition on polished single-crystal (111) silicon substrates in a plasma beam sputter deposition system. It was characterized by XTEM and RBS and flat interfaces have been found \cite{Barna}. AESD depth profiling was carried out using a dedicated device \cite{Barna2} by applying Ar$^+$ ions of energy of 1 keV and angle of incidence of $10^{\circ}$ with respect to the surface of the crystal (grazing angle of incidence). The sample was rotated during ion bombardment. 

 The depth profiles were measured as a function of the sputtering time keeping the bombarding ion current constant. A STAIB DESA 100 pre-retarded CMA with fixed energy resolution was used to record the AES spectra. The following AES peaks were detected Cu (60 eV), Cu (920 eV) and Co (656 eV, to avoid overlapping). A part of a typical depth profile is shown in Fig 1. For clarity only the copper is shown in Fig 1. To calculate the concentration the intensity of the copper Auger current was normalized to that of the pure copper. The depth scale was calculated from the known thickness of the sample \cite{Barna}.

It is generally supposed that during ion mixing the atomic movements are similar to that of the usual diffusion and thus the occurring broadening can be described by the same equations \cite{Bolse}. Accordingly in case of a bilayer system the concentration distribution formed due to ion bombardment of the interface by a given fluence can be described by the $erf$ function. The variance, $\sigma^2$, of the $erf$ function determines the extent of ion mixing.
 In many cases $\sigma^2$ is linearly dependent on the bombarding ion fluence $\Phi$. In these cases the $\sigma^2/\Phi$ ratio called as mixing rate characterizes the mixing for a given ion bombarding condition. Important
advantage of using the term mixing rate is that it can be in principle directly derived from the experiments. 

\begin{figure}[hbtp]
\begin{center}
\includegraphics*[height=5cm,width=6cm,angle=0.]{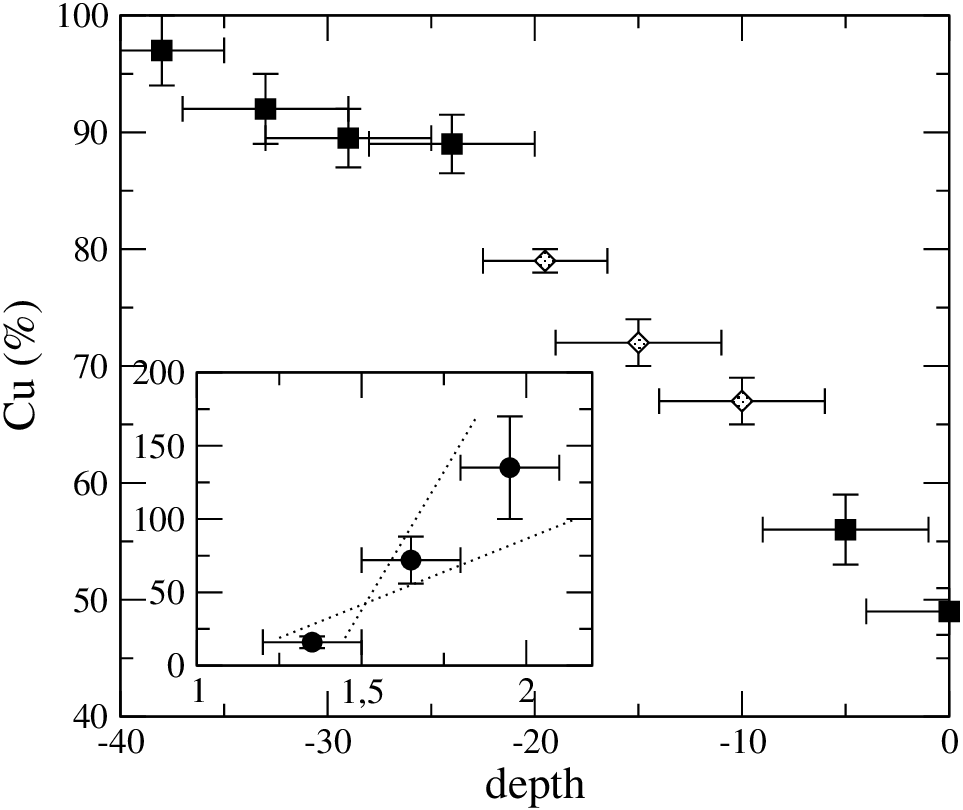}
\caption[]{
The measured depth profile: projectile Ar$^+$, ion energy 1 keV,
angle of incidence 84$^{\circ}$.
The concentration of Cu is given as a function of the removed
layer thickness (measured from the interface in $\hbox{\AA}$).
The diamonds denote the points at which $\sigma^2$ values are determined
by AESD.
{\em Inset figure}:
The broadening ($\sigma^2$) of the interface as a function of the fluence (ion/
$\hbox{\AA}^2$).
The dotted lines show the two extreme of the possible slopes (mixing rates).
}
\label{fig1}
\end{center}
\end{figure}

  In the case of AESD we measure the Auger current of the elements present. The measured Auger current depends on the in-depth distribution as follows:
$I_j=i_j(1)+i_j(2) \kappa_j+i_j(3) \kappa_j^2+...$, where $I_j$  is the measurable Auger current of element $j$,  $i_j(k)$ is the Auger current of element $j$ emitted by the atomic layer $k$, and $\kappa_j$ gives the attenuation of the Auger current crossing an atomic plane. $i_j(k) \approx X_j(k)N(k)$, where $X_j(k)$ is the concentration of element $j$ in layer $k$, while $N(k)$ is the number of atoms  of the $k$-th atomic plane. It is evident that in general from a single measured $I_j$ one cannot determine the $i_j(k)$ values. On the other hand this equation can be used to simulate the measured Auger current during the depth profiling procedure if we assume an in-depth distribution.

 We do not know any experimental measurement of the in-depth distributions formed during AESD applying low ($0.2-2$ keV) ion energy. It seems, however, that dynamic TRIM simulation can reliably be used to describe AESD \cite{Trim}. This calculation also provides the in-depth distribution during the procedure. It turns out that at the beginning of the depth profiling procedure, when the interface is still far from the surface the in-depth distribution can be approximated by the $erf$ function. For the evaluation of the
experimentally determined depth profile we will suppose that
(i) the interface has an intrinsic surface roughness, which can also be approximated by an $erf$ function of $\sigma_0$. 
(ii) ion mixing is the only process contributing to the broadening of the interface at least in the beginning part of the depth profiling, which will be studied. 
Thus for any measured copper Auger current we should find $x_0$ (the distance of the interface from the surface) and $\sigma_m$ (the measured variance) values of the erf function. Then the variance due to the ion bombardment induced mixing is $\sigma^2= (\sigma_m^2-\sigma_0^2)$.
We derived $\sigma^2$ at three depths (indicated in Fig 1.) to be $16 \pm 4$ $\hbox{\AA}^2$, $72 \pm 16$ $\hbox{\AA}^2$, and $135 \pm 35$ $\hbox{\AA}^2$. To proceed we must know the number of ions causing the broadening at the interface. In our experimental arrangement we cannot measure the ion fluence. On the other hand we can measure accurately the removed layer thickness. Taking the sputtering yield from the literature to be $Y \approx 1.2$ \cite{Barna}, we can derive the curve $\sigma^2$ vs. fluence $\Phi$ which is shown in inset Fig. 1. Inset Fig. 1 also 
\begin{figure}[hbtp]
\begin{center}
\includegraphics*[height=5cm,width=6cm,angle=0.]{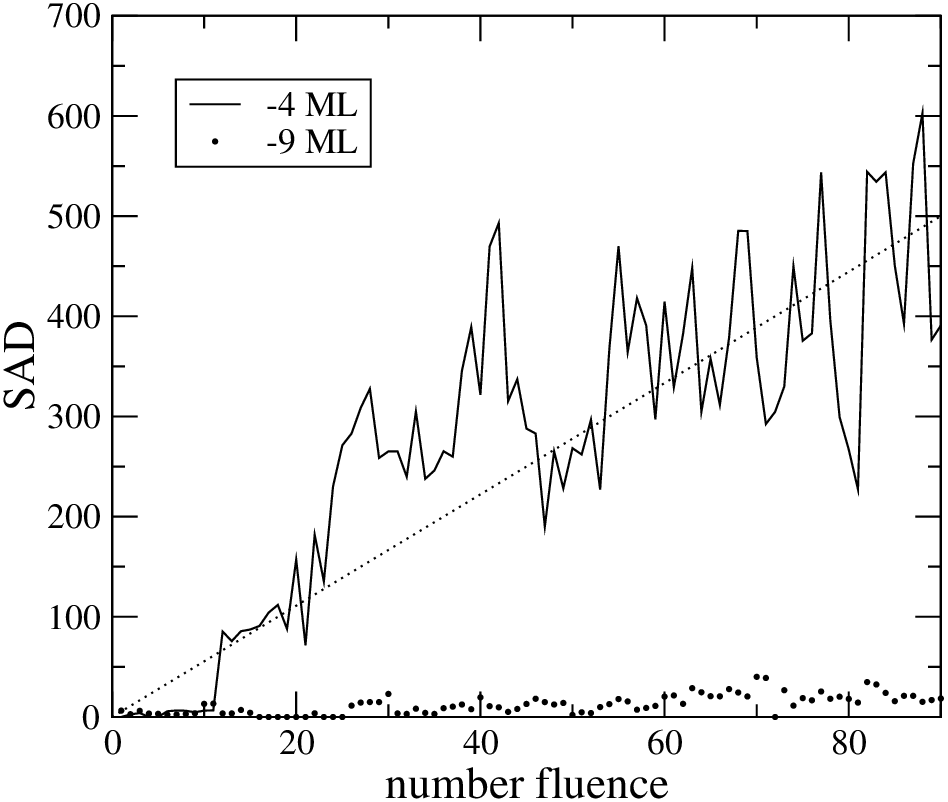}
\caption[]{
The simulated filtered $\langle R^2 \rangle$ (SAD) as a function of the number of the ion impacts (number fluence)
at 1 keV ion energy at grazing angle of incidence using simulated ion-sputtering (molecular
dynamics).
Dotted line is a linear fit to the $\langle R^2 \rangle$ curves.
$\langle R^2 \rangle$ is simulated with ion impact points at the free surface (9 ML above the interface) and also at $4$ ML above the interface.
}
\label{fig2}
\end{center}
\end{figure}
shows the two limiting slopes; thus we can derive the mixing rate being
$250 \pm 150$ $\hbox{\AA}^4$.

  Classical constant volume molecular dynamics simulations were also used to simulate the ion-solid interaction
(ion-sputtering) using the PARCAS code \cite{Nordlund_ref}.
The computer animations can be seen in our web page \cite{web}.
Further details  are given in ref. 
\cite{Nordlund_ref,Sule_PRB05,Sule_NIMB04}.
We irradiate the bilayer Cu/Co (9 monolayers, (ML) film/substrate)
with 1 keV Ar$^+$ ions repeatedly with a time interval of 5-20 ps between each of
the ion-impacts at 300 K
which we find
sufficiently long time for the termination of interdiffusion, such
as sputtering induced intermixing (ion-beam mixing) \cite{Sule_NIMB04}.
 The initial velocity direction of the
impinging atom was $10^{\circ}$ with respect to the surface of the crystal (grazing angle of incidence)
to avoid channeling directions and to simulate the conditions applied during ion-sputtering.

 To describe homo- and heteronuclear interaction of Cu and Co, the Levanov's \cite{Levanov} tight-binding
potentials are used \cite{CR}.
The cutoff radius $r_c$ is taken as the second neighbor distance.

 We randomly varied the impact position and the azimuth angle $\phi$.
In order to approach the real sputtering limit a large number of ion irradiation are
employed using automatized simulations conducted subsequently together with analyzing
the history files (movie files) in each irradiation steps.
In this article we present results up to 100 ion irradiation which we find suitable for
comparing with low to medium fluence experiments. 100 ions are randomly distributed
over a $40 \times 40$ \hbox{\AA}$^2$ area.
The size of the simulation cell is $100 \times 100 \times 75$ $\hbox{\AA}^3$ including
62000 atoms.
\begin{figure}[hbtp]
\begin{center}
\includegraphics*[height=5cm,width=6cm,angle=0.]{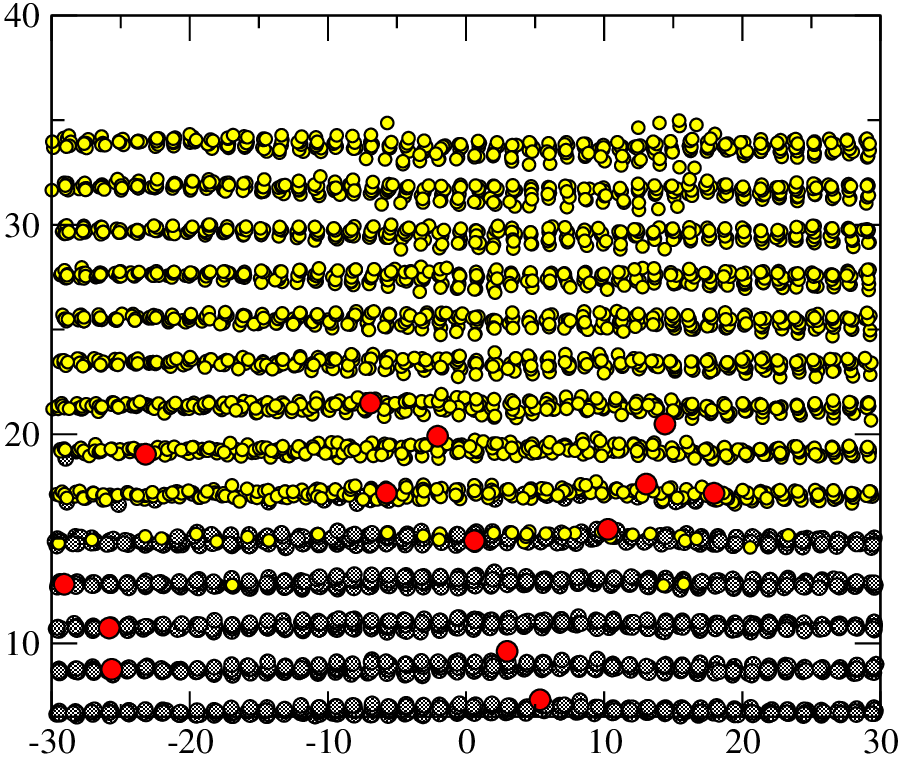}
\caption[]{
The crossectional view of the ion-sputtered Cu/Co bilayer at 1 keV
Ar$^+$ ion energy (ions are initialized at -4ML) obtained by simulated ion-sputtering. The red circles are the incorporated
ions. 
}
\label{fig3}
\end{center}
\end{figure}
  Chanelling recoils are left to move outside the cell
and in the next step these energetic and the sputtered particles are deleted.

 Unfortunately the realistic simulation of the layer-by-layer sputter-removal is beyond
the performance of the computers available. Just to give a hint of the difficulties,
the simulation of few thousands of ion impacts and subsequent relaxations should be treated.
To reduce the computational demand,
we carried out
 two types of simulations. In the first case ions are
initialized from the surface (9 ML far from the interface).
$100$ ions corresponds to $\Phi \approx 0.063$ ion/$\hbox{\AA}^2$ (the removal of
$\sim 0.35$ ML).
 If the ion is initialized e.g. 5 ML below the free surface (-4 ML far from the interface), than
we can simulate high dose experiments, when nearly 4 ML is sputter-removed.
Although, the removal of the upper 4 ML has not been done,
 we expect that this kind of an artificial setup of ion-sputtering
can simulate the real experiment.
This is because, we expect that the main source of interfacial mixing
is the progressive enhancement of energy deposition at the interface with respect
to the depth position of the ion-impacts. 

  In Fig 2 the evolution of the sum of the square atomic displacements (SAD) along the depth direction of all intermixing atoms $\langle R^2 \rangle= \sum_i^N [z_i(t)-z_i(t=0)]^2$, where $z_i(t)$ is the depth position of atom $i$ at time $t$, can be followed as a function of the 
number of ions.
Using the relation
$\sigma^2 = \langle R^2 \rangle$ it is possible to calculate the mixing rate
($k=\langle R^2 \rangle/\Phi$) from the slope of the fitted straigh line
on the $\langle R^2 \rangle$ vs. $\Phi$ curve in Fig 2.
The contribution of intermixed atoms to $\langle R^2 \rangle$ is excluded in a given layer  
 if the in-layer concentration of them is less then $\sim 5$ \%.
AESD can not measure intermixed atoms which has very low concentration
and which are below the treshold sensitivity of AESD.
In the 2nd and 3rd Co layers below the interface we find less then $5$ and
$1$ \% Cu which are below the sensitivity of AESD measurements.
Ion-sputtering in the 1st (-9 ML) and in the 5th (-4 ML) layers 
$k_{-9 ML} \approx 317$ $\hbox{\AA}^4$ and $k_{-4 ML} \approx 397$ $\hbox{\AA}^4$, are obtained, respectively.
The corresponding fluences are $\Phi \approx 0.063$ ion/$\hbox{\AA}^2$ and
$\Phi \approx 1.26$ ion/$\hbox{\AA}^2$ (the artificial removal of 4 ML corresponds to $\Phi \approx 1.2$ ion/$\hbox{\AA}^2$ using the $Y \approx 1.2$ ion/atom), respectively.
It must also be noted that Cai {\em et al.} obtained
$k \approx 400$ $\hbox{\AA}^4$ using 1 MeV Si$^+$ ions and x-rax scattering techniques
in Co/Cu multilayer \cite{Cai}.
Since the ion mixing occur at the low energy end of the cascade process, e.g.
the high energy collisional cascades split into low energy subcascades (see e.g. refs. in \cite{Sule_PRB05}),
an agreement is expected for.
We also give the measured and calculated mixing efficiencies of $\xi =k/F_D$, where
$F_D$ is the deposited ion-energy/depth.
We get $\xi
 \approx 10 \pm 2$ $\hbox{\AA}^5$/eV both by experiment and by simulations.
On the basis of this value we can characterize low-energy ion-sputtering induced intermixing in Cu/Co
as a ballistic interdiffusion process.

  The crossectional view of the ion-sputtered system can be seen in Fig 3 as obtained
by MD simulations. The interface is only weakly intermixed.
The incorporated ions are also shown in Fig 3.
The simulation can also be seen as an animation \cite{web}.

  In this Letter we have presented
that the combination of atomistic simulations with
Auger electron spectroscopy depth profiling might be a new efficient method
to depth profiling analysis of multilayered materials.
Also, the reasonably good agreement between experiment and simulations
provides us the possibility of predicting interdiffusion properties for various multilayers
for which no experimental results are available.

\section{acknowledgement}
{
\scriptsize
This work is supported by the OTKA grant F037710
from the Hungarian Academy of Sciences.
The work has been performed partly under the project
HPC-EUROPA (RII3-CT-2003-506079) with the support of
the European Community using the supercomputing
facility at CINECA in Bologna.
The help of the NKFP project of
3A/071/2004 is also acknowledged.
}

\end{document}